\documentclass[12pt]{iopart}
\usepackage{graphicx}

\begin{document}
\title{Extremely low-frequency electromagnetic fields cause DNA strand breaks in normal Vero cells}

\author{Cosmin Teodor Mihai$^1$ $^*$,
        Gabriela Vochita$^2$,
        Florin Brinza$^3$,
       Pincu Rotinberg$^2$
      }

\address{$^1$ $^*$ Department of Biology, "Alexandru Ioan Cuza" University of Iasi,\\
      Bd. Carol I, nr. 20A, 700505, Iasi, Romania}
\ead{cosmin.mihai@uaic.ro}
\address{$^2$ National Institute of Research and Development for Biological Sciences, branch Institute of Biological Research Iasi,\\
   Str. Lascar Catargi, nr. 47, 700107,Iasi, Romania}
\address{$^3$ Faculty of Physics, “Alexandru Ioan Cuza” University\\
Bd. Carol I, nr. 11, 700506, Iasi, Romania}
      
\begin{abstract}
Extremely low frequency electromagnetic fields aren't considered as a real carcinogenic agent despite the fact that some studies have showed impairment of the DNA integrity in different cells lines.\\
The aim of this study was evaluation of the late effects of a 100 Hz and 5.6 mT electromagnetic field, applied continuously or discontinuously, on the DNA integrity of Vero cells assessed by alkaline Comet assay and by cell cycle analysis. 
Normal Vero cells were exposed to extremely low frequency electromagnetic fields (100 Hz, 5.6 mT) for 45 minutes. The Comet assay and cell cycle analysis were performed 48 hours after the treatment.\\
Exposed samples presented an increase of the number of cells with high damaged DNA as compared with non-exposed cells. Quantitative evaluation of the comet assay showed a significantly ($<$0.001) increase of the tail lengths, of the quantity of DNA in tail and of Olive tail moments, respectively. \\
The analysis of the registered comet indices showed that an extremely low frequency electromagnetic field of 100 Hz and 5.6 mT had a genotoxic impact on Vero cells. Cell cycle analysis showed an increase of the frequency of the cells in S phase, proving the occurrence of single strand breaks. The most probable mechanism of induction of the registered effects is the production of different types of reactive oxygen species.
\end{abstract}

\vspace{2pc}
\maketitle

\section{Introduction}
The extremely low frequency electromagnetic fields (ELF-EMF) are omnipresent in human life, being generated by common appliances electrical conductors that cross the populated areas or the walls of houses, medical devices used in the treatment of different illness, electrical cars (used in the public transportation systems or as private cars) and electrical trains (underground or suburban electrical trains). Both in the case of the medical devices (mainly used in the physiotherapy) and of the electrical components of the cars and trains, the common generated frequency is that of 100 Hz \cite{journal-1}. Even though the patients or passengers are exposed for a short time, the deserving personnel are subject to prolonged exposure. Also, the combined treatment of 100 Hz magnetic field and X-rays has increased the survival time of hepatoma-implanted Balb/c mice as compared to magnetic field or X-rays treated groups \cite{journal-2}, suggesting the possible use in oncotherapy.\\
The exposure of different cell lines or organisms to electromagnetic fields have produced a bulk of data that  were sometimes contradictory and didn't allow a concise and clear conclusion about the effects of electromagnetic fields on biological systems \cite{journal-3, journal-4}. \\
The International Agency for Research on Cancer (IARC) has evaluated the scientific data and has classified ELF magnetic fields as being "possibly carcinogenic" to human \cite{journal-5}. \\
The worrying but contradictory epidemiologic or experimental data about the possible genotoxic effects of this kind of electromagnetic fields, suggesting the possible carcinogenetic effect, requires an enrichment of the pool of experimental information and the deciphering of its action mechanisms. \\
Carcinogenic processes have three developmental stages: initiation, promotion and progression. The first stage, tumour initiation, begins when the DNA in a cell or population of cells is damaged by exposure to exogenous or endogenous carcinogens. If this damage is not repaired, it can lead to genetic mutations. The responsiveness of the mutated cells to their microenvironment can be altered and may give them a growth advantage relative to normal cells \cite{journal-6, journal-7, journal-8, journal-9}. \\
The possible carcinogenetic effect of the low frequency and intensity electromagnetic fields are still under debate, the data being controversial. Studies in this field suggested that exposure to low frequency and intensity electromagnetic fields could alter the DNA integrity, which could trigger the initiation of carcinogenetic processes or could accelerate the development or spreading of already present cancers \cite{journal-10, journal-11}. Also, it was suggested that chronic exposure to the ELF could be involved in the development of some neurodegenerative diseases by production of reactive oxygen species \cite{journal-12}. \\
Contrary, other researches identified no effects on the integrity of the DNA in the conditions of exposure to the electromagnetic fields \cite{journal-13, journal-14, journal-15, journal-16, journal-17}. \\
The aim of the study was to test if, in effect, there are any differences between continuous or discontinuous extremely low frequency electromagnetic fields on the DNA integrity of normal Vero cells, in order to evaluate the possible disruptions that could lead to mutagenicity. The evaluation of the effects of the extremely low frequency electromagnetic fields was assessed by alkaline comet assay and cell cycle analysis.\\

\section{Materials and methods}
\subsection{Cell cultures}Vero cells (ECACC 88020401) are adherent to substratum with a fibroblast-like morphology. The cells were cultivated in a DMEM medium (Dulbeco's Modified Eagle’s Medium, Biochrom AG, Germany, FG 0415), supplemented with 2.0\% foetal bovine serum (Sigma, Germany, F9665) and 100 $\mu$g/mL streptomycin (Biochrom AG, Germany, A 331-26), 100 IU/mL penicillin (Biochrom AG, Germany, A 321-44). The cell cultures were seeded at a density of 5 x 10$^5$ cells in 25 cm$^2$ flask (TPP Techno Plastic Products AG, Trasadingen, Switzerland) and maintained in a CO$_2$ incubator (Binder CB 150, Tuttlingen, Germany), at 37$^o$C.  When the cells reached confluence in the monolayer stage, the cultures were divided into control and electromagnetic treated cell cultures. \\
\subsection{ELF-EMF exposure setup }
\begin{figure}[h]
\centering
\includegraphics[scale=1.2]{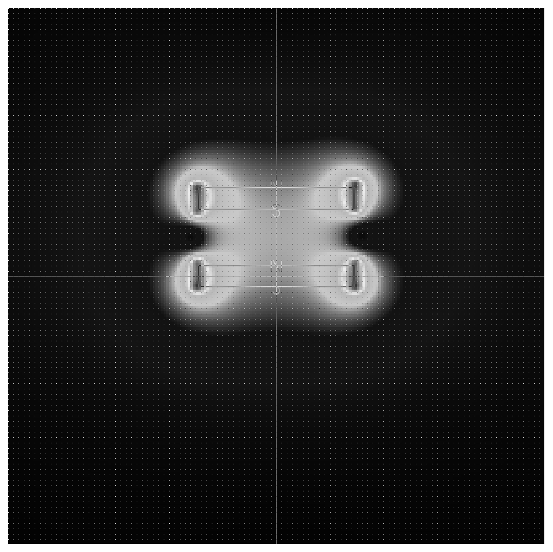}
\qquad
\includegraphics[scale=1]{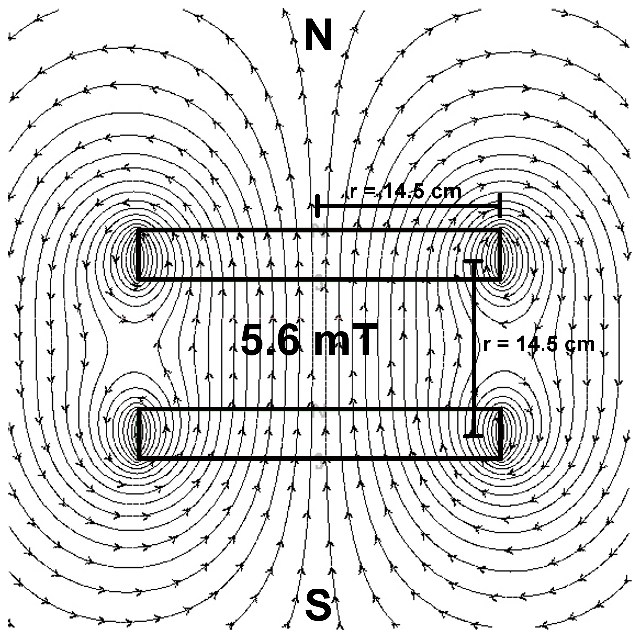}
\caption{\label{fig.1}{Magnetic flux density (left image) and flux lines distribution (right image) in the area of exposure of the cells when the Helmholtz coils were powered with a direct current with the frequency of 100 Hz and an intensity of 2.0 Amperes.}}
\end{figure}

The setup for electromagnetic exposure consisted in a Helmholtz pair of coils connected in parallel to a magnetodiaflux (IBF, Romania) device, that generates a pulse electromagnetic field (PEMF) having a frequency of 100 pulses/second. \\
The 29 cm diameter coils were made of copper wires with 620 turns. The coils were set at a distance equal with their radius (14.5 cm) which assured a central homogeneous magnetic field (5.6 mT). The unpowered coils presented a magnetic field with a 0.021 mT intensity, similar to the registered magnetic field background. 
The magnetodiaflux device delivered to the two coils a pulsating direct current (PDC), obtained by converting and rectifying the 220 V/50 Hz alternative current. The PDC had a 100 Hz frequency and the peak voltage variation, measured with an oscilloscope (Tektronix, Guernsey, Channel Islands), was of 42 volts and 2.0 Ampers. \\
The coils were housed in the cell incubator, the temperature being constant and uniformly distributed all the time of the exposure (37$\pm$0.2$^o$C), as monitored by a thermocouple thermometer (Hanna Instruments, Italy). 
The homogeneity of the magnetic field produced by the coils in the area of exposure is shown in the figure \ref{fig.1}, image being generated by Vizimag ver.3.193 software (\copyright  J. Beeteson 1999-2009), using the provided characteristics of the coils and current. \\

\subsubsection{Magnetic flux density measurement}
The magnetic flux density measurement inside of Helmholtz coils system was performed using LakeShore 421 Gaussmeter, having valid NIST certificates.\\
Both axial and transverse probes were used for the measurements. Before the measurement process, each probe was calibrated in a zero-field chamber. For the characterization of the uniformity in the samples volume, experimental measurements and software simulation were used. The measurements were performed in equidistantly distributed points inside of coils. In each point, axial (using axial probe) and radial (using transverse probe) values of the magnetic flux density were tested. The experimentally detected values were compared to the calculated ones.\\
The field lines distribution and the magnetic induction values in the centre of the coils system were calculated using a Vizimag ver.3.193 software. In Figure \ref{fig.1}, the field lines distribution and the field lines density for our coils are presented. The experimental measured values of the magnetic flux density are in good accordance with calculated ones. The conclusion is that the whole volume of cell culture container is subjected to a uniform magnetic induction value. The low values of magnetic susceptibility of the biological samples ensure a uniform value of the magnetic field inside the container volume.\\
\subsection{Treatment of the cells }
The cell culture flasks with the cells were placed in the central region of the Helmholtz coils, perpendicular to the magnetic field lines. The cells were exposed once for 45 minutes and were returned to the incubator after the treatment. The exposure of the cells was performed in a continuous (cEMF, permanent exposure to the magnetic field during the 45 minutes of the treatment) or in a discontinuous manner (dcEMF, cycles of one second on and three seconds off). Sham-exposed cells were put into the same experimental conditions as the treated samples but without energizing the coils, the EMF background remaining practically unchanged. \\
To asses the late effects of the exposure to ELF-EMF, the culture flasks were returned to the incubator and maintained for another 48 hours after the evaluation of DNA integrity by alkaline Comet assay and cell cycle analysis were performed. \\
\subsection{Comet assay analysis}
We followed the technique described by Ostling and Johanson \cite{journal-18} with minor modifications by Singh et al. \cite{journal-19}, as presented in \cite{journal-20}. \\
The 50 µL of ELF-exposed and the sham-exposed cells (20,000 cells) were mixed with 150 $\mu$L of low-melting agarose (0.8\%, 37$^o$C), and this cell suspension was pipetted onto 1\% normal-melting agarose pre-coated slides, spread with a cover slip, and kept on a cold flat tray for approximately 10 minutes to solidify. \\
 The slides were then immersed in freshly prepared cold lysis solution (2.5 mol/l NaCl, 100 mmol/l Na$_2$EDTA, 10 mmol/l Tris, pH 10, 1\% sodium sarcosinate, 1\% Triton X-100, 10\% DMSO, pH 10) and lysed overnight at 4$^o$C. Subsequently, the slides were drained and placed in a horizontal gel electrophoresis tank, side by side and very close to the anode. The tank was filled with fresh electrophoresis buffer (1 mmol/l Na$_2$EDTA, 300 mmol/l NaOH, pH$>$13 to a level approximately 0.4 cm above the slides. The slides were left in the solution for 40 minutes, to allow equilibration and unwinding of the DNA before electrophoresis. \\
The electrophoresis was conducted at 25 V, 300 mA, 4$^o$C, 20 min, field strength 0.8 V/cm. All steps were performed under dimmed light to prevent the occurrence of additional DNA damage. After electrophoresis, the slides were washed three times with Tris buffer (0.4 mol/l Tris, pH 7.5), to be neutralized, then air-dried and stored until needed for analysis. Comets were visualized by ethidium bromide staining (20 $\mu$g/ml, 30 s) and examined at 200 magnification with a fluorescence microscope (Nikon Eclipse 600, Nikon corp., Japan).\\
\subsection{Image analysis of comets}
The image analysis was performed with CASP software (CASP or Comet Assay Software Project, http://www.casp.sourceforge.net). The evaluation of comets was done by tail length, \% content of DNA in tail and by calculating tail and Olive moment \cite{journal-21, journal-22}.\\
\subsection{Cell cycle analysis}
After the electromagnetic treatment cells were harvested from the surface of culture flasks by trypsinization, they were resuspended in a complete medium and then pelleted by centrifugation. The cells were washed twice in cold PBS. The cell pellet was resuspended in NIM-DAPI (Beckman Coulter, USA) and were stained overnight at 4$^o$C. For every control and treated sample, 20.000 cells were measured on flowcytometer, using a 100 W mercury arc lamp, a 355/37 exciter and a 460 BP filter for the collection of fluorescence and linear amplification. \\
\subsection{Statistical analysis}
All of the experiments were carried out with at least three independent repetitions and all data were expressed as the mean value and standard error of mean (SEM). The statistical analysis was performed using Student's “t” test and the differences were expressed as significant at the level of p$<$0.05.

\section{Results}
The qualitative analysis of the cellular damage determined by electromagnetic fields was evaluated by the extent of the damage and graded according to \cite{journal-20}, the results being presented in table \ref{tab1}.\\

\begin{table}[!h]
\caption{Absolute frequencies of Comet types found in control group and in samples treated with cEMF and dcEMF (n = 400 cells) for 45 minutes once and evaluated by Comet assay after 48 hours from the treatment.\label{tab1}}
\begin{tabular}{cccccc}
\hline
 &  {Control}  &  \multicolumn{2}{c}{cEMF 100 Hz}	& \multicolumn{2}{c}{dcEMF 100 Hz}\\
 \cline{2-6}
&	\%Mean$\pm$SEM &  \%Mean$\pm$SEM & p & \%Mean$\pm$SEM & p\\ 
     \hline
    A ($<$5\%)	& 59.71$\pm$0.10	 & 55.83$\pm$0.08 & $<$0.002	& 66.11$\pm$0.07&	NS\\
B (5 $-$ 20\%)&	24.82$\pm$0.49&	21.80$\pm$0.34&	$<$0.001&	8.61$\pm$0.81&	$<$0.001\\
C (20 $-$ 40\%)	&10.79$\pm$1.01&	 9.59$\pm$0.83&	 $<$0.001	& 8.61$\pm$1.00 &	$<$0.001\\
D (40 $-$ 95\%)	& 4.68$\pm$1.81 & 	12.22$\pm$1.29&	$<$0.02	&16.11$\pm$1.43	& $<$0.05\\
E ($>$95\%)	&0.00$\pm$0.00&	0.56$\pm$0.01&	$<$0.001&	0.56$\pm$0.00&	$<$0.001 \\
\hline
 \end{tabular}
\paragraph{}
Errors indicate the standard error of the mean (SEM) for n=3 independent experiments.
\end{table}

\begin{figure}[h]
\centering
\includegraphics[width=2.5in]{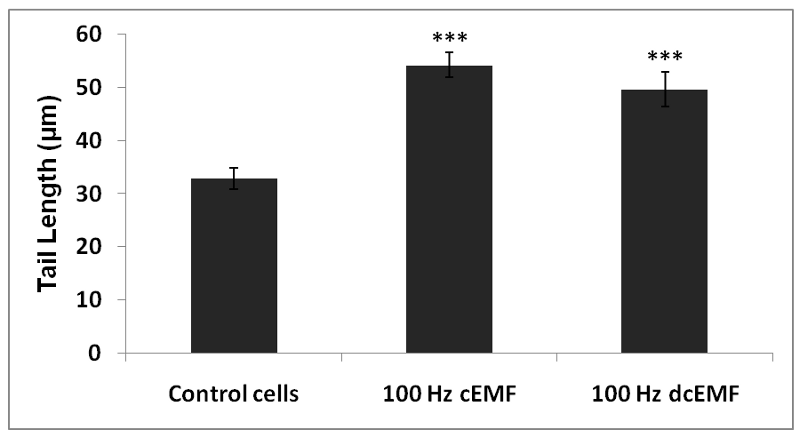}
\qquad
\includegraphics[width=2.5in]{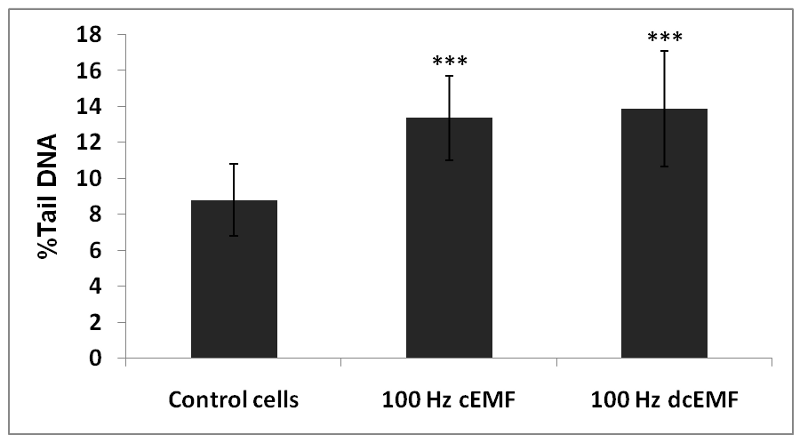}
\caption{\label{fig.2}{Tail length (left figure) and \% content in DNA (right figure) of the tail determined in Vero cells at 48 hours after the exposure to extremely low frequency electromagnetic fields (100 Hz, 5.6 mT, 45 minutes). *** = $<$0.001  }}
\end{figure}

The number of the cells that presented DNA damage higher than 40\% was greater in the case of the cells subjected to the electromagnetic field treatment, compared to the control group. dcEMF had a more negative impact upon the integrity of the genetic material of Vero cells as compared to cEMF. In both cases, the differences were significant ($<$0.05).\\
The quantitative evaluation was based on four parameters: tail length (TL), \%DNA in tail (TD), tail moment (TM) and Olive tail moment (OTM).\\
The tail length gives us information about the dimension of DNA fragments (smaller fragments, longer tail) and it is expected to be proportional to the extent of DNA damage. As shown in figure \ref{fig.2}, the cEMF and dcEMF determined an increase of the tails in treated cells, the values of tail length being 1.65 (54.20  $\mu$m$\pm$ 2.34), respectively 1.51 (49.63  $\mu$m$\pm$ 3.23) times bigger than in the case of the control group (32.81  $\mu$m$\pm$ 2.01). The differences between the treated and the control samples, evaluated by t test, were found to be statistically significant ($<$0.001). The second parameter taken into consideration in the evaluation of the effects of the electromagnetic fields upon the integrity of the DNA of Vero cells was \%DNA in tail, which provides data about the damaged DNA content in individual cells, measured as the total intensity of ethidium bromide in every cell subjected to the electrophoresis in alkaline conditions. The \% DNA in tail increased significantly ($<$0.001) in the cells exposed to the cEMF 100 Hz (13.37\%$\pm$0.88) and respectively to dcEMF 100 Hz (13.87\%$\pm$1.17), when compared to the control group (8.8\%$\pm$0.78), the values being of 1.52 and 1.58, respectively, times bigger than the control value. 

\begin{table}[!h]
\caption{Impact of the extremely low frequency electromagnetic field (100 Hz, 5.6 mT, 45 minutes once) on comet assay indices (tail moment and Olive tail moment) in Vero cell lines, after 48 hours from the moment of the treatment.\label{tab2}}
\begin{tabular}{cccccc}
\hline
& \multicolumn{2}{c}{Tail Moment} &	& \multicolumn{2}{c}{Olive TailMoment}\\
& Mean($\mu$m) $\pm$ SEM	& p & 	& Mean ($\mu$m) $\pm$ SEM &	 p\\
\cline{2-3} \cline{5-6}
Control cells&	8.26$\pm$1.03&	- &	&6.58$\pm$0.66&	-\\
100 Hz cEMF &	17.40$\pm$1.54	& $<$0.001 & 	&12.30$\pm$0.91	& $<$0.001\\
100 Hz dcEMF	& 20.77$\pm$2.16&	$<$0.001 &	& 12.70$\pm$1.20	& $<0$.001\\
\hline
 \end{tabular}
\paragraph{}
Errors indicate the standard error of the mean (SEM) for n=3 independent experiments.
\end{table}

The calculated parameters tail moment and the Olive tail moment correlate the extent of the tail with the quantity of DNA present in tail. \\
The tail moment represents the product of the tail length (also called the tail extent) and the percentage of DNA in the tail. EMF treatment determined an increase of TM (17.40 $\mu$m$\pm$ 1.54 for cEMF and 20.77  $\mu$m$\pm$ 2.16 for dcEMF) as compared to the control group (8.26  $\mu$m$\pm$ 1.03).\\
The olive tail moment is calculated as a product of two factors: the percentage of DNA in the tail and the distance between the intensity centroids of the head and the tail along the x-axis of the comet. OTM incorporates a measure of both the smallest detectable size of migrating DNA (reflected in the comet tail length) and the number of relaxed/broken pieces (represented by the intensity of DNA in the tail). \\
As in the case of the other parameters, OTM has increased both in the case of cEMF (12.30 $\mu$m$\pm$ 0.91) and of dcEMF (12.70  $\mu$m$\pm$ 1.20) when compared with the control group (6.58 $\mu$m$\pm$ 0.66), the calculated differences being statistically significant. The registered values of OTM for cEMF and dcEMF were similar, with a small increase in the case of dcEMF. \\

Cell cycle analysis

\begin{table}[!h]
\caption{Impact of the extremely low frequency electromagnetic field (100 Hz, 5.6 mT, 45 minutes once) on comet assay indices (tail moment and Olive tail moment) in Vero cell lines, after 48 hours from the moment of the treatment.\label{tab3}}
\centering
\begin{tabular}{llll}
\hline
& \multicolumn{3}{l}{Cell cycle distribution (\%)}\\
\cline{2-4}
	& G0/G1 stage	& S stage	& G2/M stage\\ 	
Control cells	& 74.94	& 6.01	& 18.86\\
cEMF 	& 58.74 * (-21.62\%) &	22.42 *** (+273.04\%)	& 18.68 (-0.95\%) \\
dcEMF	& 68.00 (-9.26\%)	& 14.45 *** (+140.43\%)	& 17.31 (-8.22\%)\\
\hline
 \end{tabular}
\paragraph[l]{}
\flushleft Data are means of three experiments, SD being $>$15\%. \\
The CV of the peaks were $>$ 8\%. \\
 $^*$ p$<$0.05 ; $^***$ p$<$0.001 by Student$'$s t test
\end{table}

Compared to controls, ELF-EMF-exposure caused a blockage of the cells in the S stage of the cell cycle. After 48 hours, the highest percentage of cells blocked in the S phase was registered in the case of cEMF followed by dcEMF, as it can be seen in table \ref{tab3}.
\section{Discussions}
The general assumption is that extremely low frequency electromagnetic fields aren’t genotoxic and do not affect the integrity of the DNA molecule, but the scientific data are still controversial and supplementary evidence is necessary to consider this physical agent as a real carcinogenic factor. \\
Other aspects that need to be clarified are the differences either in effect or in the magnitude of the effect, in respect to the use of continuous and intermittent electromagnetic fields. Ivancsits \cite{journal-20, journal-23, journal-24} reported that intermittent electromagnetic fields caused DNA damage of the human diploid fibroblasts, while continuous electromagnetic field did not alter the DNA integrity. Starting from this experiment, Crumpton \cite{journal-25} argued that supplementary data, obtained in independent experiments, concerning the biological effects of low frequency electromagnetic, is necessary, and the link between electromagnetic fields and initiation of the carcinogenesis processes has to be proved.\\
The current study was designed to establish whether 100 Hz extremely low-frequency electromagnetic fields affect the DNA integrity in the normal Vero cell line, the DNA damage being assed 48 hours after the electromagnetic exposure. In addition, we have investigated if there are any significant differences between the manner of exposure of 100 Hz ELF (continuous or discontinuous) and the magnitude of the effects. 
The use of the COMET assay is a very valuable tool in the identification of possible genotoxic effects of different agents and could offer rapid and solid evidence about the possible impairments of the DNA integrity. Also, this test was used extensively in a study concerning the effect of different types of electromagnetic fields upon the DNA integrity \cite{journal-4}.\\
The qualitative analysis of the comet assay showed a reduction of the frequency of cells with broken DNA between 5-40\%, but a significant increase in frequency of the cells with major impairments of the DNA. The late determination of the cells status (48 hours after the treatment) showed that the exposure to the electromagnetic field determined a spectrum of consequences upon the integrity of genetic material, such as damages which activated internal control systems of the cell and the restoration of the normal state of DNA molecule and alterations that couldn’t be corrected or needing a longer time to be repaired. The activation of the repairing systems allowed the correction of the errors with a higher efficiency than in the unexposed cells. Nevertheless, the increased frequency of the cells with a high degree of damaged DNA indicates heterogeneity of cellular response to the electromagnetic field, depending on the cellular state in the moment of exposure. The cell cycle analysis of the exposed cells revealed a blockage of the cells in the S phase of the cell cycle, suggesting that the 100 Hz ELF could impair the synthetic and correction mechanisms, as was suggested by \cite{journal-26}. In addition, the stop of the cells in S phase of the cell cycle suggests the formation of the single strand breaks or collapse of the DNA replication forks during the S phase, under the action of the electromagnetic treatment \cite{journal-27} \\
It is not clear whether the action of the electromagnetic fields is due to a direct or indirect effect, mediated by other mechanisms, as the production of reactive species of oxygen. Some results support redox-mediated ELF-EMF biological effects, as a positive modulation of antioxidant defences was observed, as well as a shift of cellular environment towards a more reduced state \cite{journal-28}. \\
Our data correlates with those obtained by Wolf et al. \cite{journal-29}, which signalled that ELF of 50 Hz determined a transient blockage of the cells in the S phase of the cell cycle and suggested the implication of the presence of reactive species of oxygen.  \\
The data obtained by qualitative evaluation of the cells assessed by comet assay were reinforced by quantitative appreciation of the electromagnetic fields effects. All four parameters determined by us showed an increase in the tail length and the quantity of fragmented DNA in the tail, tail moment and Olive tail moment. 
The overall evaluation of our findings is that extremely low frequency electromagnetic field acts as a moderate damaging agent. They are in accordance with other studies which suggested the mild oxidative effects of extremely low frequency electromagnetic fields responsible for the DNA damage \cite{journal-30}.\\
Albeit it was signalled that the intermittent exposure showed a stronger effect in the comet assay than continuous exposure \cite{journal-20, journal-31}, our results did not find significant differences between the two ways of exposure. The lack of any significant differences between continuous and discontinuous exposure of the cells to ELF-EMF, in our experimental setup, resides in the triggering changes in the metabolic pattern of the used cells, dependent by intensity of the magnetic field and not by the frequency. The modifications induced by ELF-EMF to DNA and their persistence are, most probably, the result of the appearance of the reactive species of oxygen and the prolongation of their lifetime extension under the influence of the magnetic field, as documented by \cite{journal-32}. \\
The persistence of the errors, even 48 hours after the exposure, indicates a persistence of reactive oxygen species, the perturbation of the cellular apparatus implied in the verification and repairing of the DNA errors and the occurrence of the SSBs in exposed cells.\\

\paragraph{Acknowledgements}
This study was possible with financial support from the Sectoral Operational Programme for Human Resources Development, project “Developing the innovation capacity and improving the impact of research through post-doctoral programmes”, POSDRU/89/1.5/S/49944


\section*{References}
\begin{thebibliography}{10}
\bibitem{journal-1} 	Tell R A, Sias G, Smith J, Sahl J, Kavet R. ELF magnetic fields in electric and gasoline-powered vehicles. Bioelectromagnetics. 2012 Apr 24;(March):1-6. 
\bibitem{journal-2} 	Wen J, Jiang S, Chen B. The effect of 100 Hz magnetic field combined with X-ray on hepatoma-implanted mice. Bioelectromagnetics. 2011 May;32(4):322-4. 
\bibitem{journal-3} 	Vijayalaxmi, Obe G. Controversial cytogenetic observations in mammalian somatic cells exposed to extremely low frequency electromagnetic radiation: a review and future research recommendations. Bioelectromagnetics. 2005 Jul;26(5):412-30. 
\bibitem{journal-4} 	Phillips JL, Singh NP, Lai H. Electromagnetic fields and DNA damage. Pathophysiology : the official journal of the International Society for Pathophysiology / ISP. 2009 Aug;16(2-3):79-88. 
\bibitem{journal-5} 	IARC Working Group on the Evaluation of Carcinogenic Risks to Humans. INTERNATIONAL AGENCY FOR RESEARCH ON CANCER IARC MONOGRAPHS ON THE EVALUATION OF CARCINOGENIC RISKS TO HUMANS. Iarc Monographs On The Evaluation Of Carcinogenic Risks To Humans. IARCPress, Lyon, France; 2002;80:27, 338. 
\bibitem{journal-6} 	Hanahan D, Weinberg RA. Hallmarks of cancer: the next generation. Cell. Elsevier Inc.; 2011;144(5):646-74. 
\bibitem{journal-7} 	Hursting SD, Slaga TJ, Fischer SM, DiGiovanni J, Phang JM. Mechanism-based cancer prevention approaches: targets, examples, and the use of transgenic mice. Journal of the National Cancer Institute. 1999 Feb 3;91(3):215-25. 
\bibitem{journal-8} 	Devi P. Basics of carcinogenesis. Health Administrator. 2004;XVII(1):16–24. 
\bibitem{journal-9} 	Khan N, Afaq F, Mukhtar H. Apoptosis by dietary factors: the suicide solution for delaying cancer growth. Carcinogenesis. 2007 Feb;28(2):233-9. 
\bibitem{journal-10} 	Blumenthal NC, Ricci J, Breger L, Zychlinsky a, Solomon H, Chen GG, et al. Effects of low-intensity AC and/or DC electromagnetic fields on cell attachment and induction of apoptosis. Bioelectromagnetics. 1997 Jan;18(3):264-72. 
\bibitem{journal-11} 	Mairs RJ, Hughes K, Fitzsimmons S, Prise KM, Livingstone A, Wilson L, et al. Microsatellite analysis for determination of the mutagenicity of extremely low-frequency electromagnetic fields and ionising radiation in vitro. Mutation research. 2007 Jan 10;626(1-2):34-41. 
\bibitem{journal-12} 	Lai H, Singh NP. Magnetic-Field-Induced DNA Strand Breaks in Brain Cells of the Rat. Environmental Health Perspectives. 2004 Jan 27;112(6):687-94. 
\bibitem{journal-13} 	Tateno H, Iijima S, Nakanishi Y, Kamiguchi Y, Asaka a. No induction of chromosome aberrations in human spermatozoa exposed to extremely low frequency electromagnetic fields. Mutation research. 1998 May 11;414(1-3):31-5. 
\bibitem{journal-14} 	Olsson G, Belyaev IY, Helleday T, Harms-Ringdahl M. ELF magnetic field affects proliferation of SPD8/V79 Chinese hamster cells but does not interact with intrachromosomal recombination. Mutation research. 2001 Jun 27;493(1-2):55-66. 
\bibitem{journal-15} 	Chemeris NK, Gapeyev AB, Sirota NP, Gudkova OY, Kornienko N V, Tankanag A V, et al. DNA damage in frog erythrocytes after in vitro exposure to a high peak-power pulsed electromagnetic field. Mutation research. 2004 Mar 14;558(1-2):27-34. 
\bibitem{journal-16} 	Testa a, Cordelli E, Stronati L, Marino C, Lovisolo G a, Fresegna a M, et al. Evaluation of genotoxic effect of low level 50 Hz magnetic fields on human blood cells using different cytogenetic assays. Bioelectromagnetics. 2004 Dec;25(8):613-9. 
\bibitem{journal-17} 	Zhijian C, Xiaoxue L, Yezhen L, Shijie C, Lifen J, Jianlin L, et al. Impact of 1.8-GHz radiofrequency radiation (RFR) on DNA damage and repair induced by doxorubicin in human B-cell lymphoblastoid cells. Mutation research. 2010 Jan;695(1-2):16-21. 
\bibitem{journal-18} 	Ostling O, Johanson KJ. Microelectrophoretic study of radiation-induced DNA damages in individual mammalian cells. Biochemical and Biophysical Research Communications. 1984;123(1):291-8. 
\bibitem{journal-19} 	Singh NP, McCoy MT, Tice RR, Schneider EL. A simple technique for quantitation of low levels of DNA damage in individual cells. Experimental Cell Research. Laboratory of Molecular Genetics, National Institute on Aging, Baltimore, Maryland 21224.; 1988;175(1):184-91. 
\bibitem{journal-20} 	Ivancsits S, Diem E, Jahn O, Rüdiger HW. Intermittent extremely low frequency electromagnetic fields cause DNA damage in a dose-dependent way. International archives of occupational and environmental health. 2003 Jul;76(6):431-6. 
\bibitem{journal-21} 	Vilhar B. Help$!$ There is a comet in my computer! [Internet]. 2004. p. 49. Available from: $http://botanika.biologija.org/exp/comet/comet$\_$guide01.pdf$
\bibitem{journal-22} 	Kumaravel TS, Vilhar B, Faux SP, Jha AN. Comet Assay measurements: a perspective. Cell biology and toxicology. 2009 Mar;25(1):53-64. 
\bibitem{journal-23} 	Diem E, Ivancsits S, R\"{u}diger HW. BASAL LEVELS OF DNA STRAND BREAKS IN HUMAN LEUKOCYTES DETERMINED BY COMET ASSAY. Journal of Toxicology and Environmental Health, Part A. Taylor \& Francis; 2002 May 10;65(9):641-8. 
\bibitem{journal-24} 	Ivancsits S, Pilger A, Diem E, Jahn O, R\"{u}diger HW. Cell type-specific genotoxic effects of intermittent extremely low-frequency electromagnetic fields. Mutation research. 2005 Jun 6;583(2):184-8. 
\bibitem{journal-25} 	Crumpton MJ, Collins AR. Are environmental electromagnetic fields genotoxic? DNA repair. 2004 Oct 5;3(10):1385-7. 
\bibitem{journal-26} 	Miyakoshi J, Yoshida M, Shibuya K, Hiraoka M. Exposure to strong magnetic fields at power frequency potentiates X-ray-induced DNA strand breaks. J Radiat Res. Department of Radiation Genetics, Graduate School of Medicine, Kyoto University, Yoshida-Konoe-cho, Sakyo-ku, Kyoto 606-8501, Japan. miyakosh@mfour.med.kyoto-u.ac.jp; 2000;41(3):293-302. 
\bibitem{journal-27} 	Caldecott KW. Single-strand break repair and genetic disease. Nature reviews. Genetics. 2008 Aug;9(8):619-31. 
\bibitem{journal-28} 	Falone S, Grossi MR, Cinque B, D$’$Angelo B, Tettamanti E, Cimini A, et al. Fifty hertz extremely low-frequency electromagnetic field causes changes in redox and differentiative status in neuroblastoma cells. The international journal of biochemistry \& cell biology. 2007 Jan;39(11):2093-106. 
\bibitem{journal-29} 	Wolf FI, Torsello A, Tedesco B, Fasanella S, Boninsegna A, D'Ascenzo M, et al. 50-Hz extremely low frequency electromagnetic fields enhance cell proliferation and DNA damage: possible involvement of a redox mechanism. Biochimica et biophysica acta. 2005 Mar 22;1743(1-2):120-9. 
\bibitem{journal-30} 	Bu\l dak RJ, Polaniak R, Bu\l dak L, Zwirska-Korczala K, Skonieczna M, Monsiol A, et al. Short-term exposure to 50 Hz ELF-EMF alters the cisplatin-induced oxidative response in AT478 murine squamous cell carcinoma cells. Bioelectromagnetics. 2012 Apr 25;(October 2011). 
\bibitem{journal-31} 	Diem E, Schwarz C, Adlkofer F, Jahn O, Rüdiger H. Non-thermal DNA breakage by mobile-phone radiation (1800 MHz) in human fibroblasts and in transformed GFSH-R17 rat granulosa cells in vitro. Mutation research. 2005 Jun 6;583(2):178-83. 
\bibitem{journal-32} 	Simk\'{o} M, Mattsson M-O. Extremely low frequency electromagnetic fields as effectors of cellular responses in vitro: possible immune cell activation. Journal of Cellular Biochemistry. 2004;93(1):83-92. 
\end{thebibliography}
\end{document}